\documentclass[%
 aip,
 amsmath,amssymb,
preprint,%
]{revtex4-1}
\usepackage{longtable}
\usepackage{CJKutf8}
\usepackage{multirow}
\usepackage{graphicx}
\usepackage{dcolumn}
\usepackage{bm}
\usepackage{lineno}
\usepackage[utf8]{inputenc}

\usepackage[T1]{fontenc}
\usepackage{mathptmx}
\usepackage[final]{changes}
\usepackage{todonotes}

\begin{document}

\linenumbers
\begin{CJK*}{UTF8}{}
\CJKfamily{gbsn}


\title[The LOG-EXP formula of the wall]{A single formula for the law of the wall and its application to wall-modelled large-eddy simulation}

\author{Fengshun Zhang {\color{black}(张风顺)}}

\affiliation{ 
The State Key Laboratory of Nonlinear Mechanics, Institute of Mechanics, Chinese Academy of Sciences, Beijing 100190, China 
}%

\affiliation{ 
Department of Mechanics and Engineering Science, Lanzhou University, Lanzhou 730000, China 
}%

\author{Zhideng Zhou {\color{black}(周志登)}}

\author{Xiaolei Yang {\color{black}(杨晓雷)}}
\email{xyang@imech.ac.cn}

\affiliation{ 
The State Key Laboratory of Nonlinear Mechanics, Institute of Mechanics, Chinese Academy of Sciences, Beijing 100190, China 
}%

\affiliation{ 
School of Engineering Sciences, University of Chinese Academy of Sciences, Beijing 100049, China 
}%

\author{Huan Zhang {\color{black}(张欢)}}

\affiliation{ 
Department of Mechanics and Engineering Science, Lanzhou University, Lanzhou 730000, China 
}%

\date{\today}

\begin{abstract}
    In this work, we propose a single formula for the law of the wall, which is dubbed as the logarithmic-exponential (LOG-EXP) formula, for predicting the mean velocity profile in different regions near the wall. And then a feedforward neural network (FNN), whose inputs and training data are based on this new formula, is trained for the wall-modelled large-eddy simulation (WMLES) of turbulent channel flows. The direct numerical simulation (DNS) data of turbulent channel flows is used to evaluate the performance of both the formula and the FNN. Compared with the Werner-Wengle (WW) model for the WMLES, a better performance of the FNN for the WMLES is observed for predicting the Reynolds stresses.     
\end{abstract}

\maketitle

\end{CJK*}
%
%
\section{Introduction}
The law of the wall is one of the cornerstones in wall-bounded turbulent flows~\cite{Pope, davidson2015turbulence}. Different formulae have been proposed in the literature for the law of the wall. Roughly, they can be divided into two groups, i.e., the piecewise formulae and the single formulae. In the early year, the piecewise function was developed to describe the dynamics in different near-wall regions of the inner layer (e.g., for turbulent channel flows, it is located in the range of $0\leq y\leq0.1\delta$, where y denotes the wall-normal direction and $\delta$ is the half-height of the channel)~\cite{Prandtl,Taylor,Karman,Deissler,Rannie,Breuer,Werner,Inagaki}, which includes the viscous sublayer, the buffer layer and the logarithmic layer~\cite{Pope}. The most widely used law of the wall describes the viscous sublayer using the linear profile, i.e., 
\begin{equation}\label{eq:linear}
    U^{+} = y^{+},
\end{equation}
where $U^+=U/u_{\tau}$, $y^{+}=y/\delta_v$ with the friction velocity $u_{\tau} = \sqrt{\tau_w/\rho}$ (where $\tau_w$ is the wall shear stress and $\rho$ is the fluid density) and the viscous scale $\delta_v = \nu/u_{\tau}$ ($\nu$ is the kinematic viscosity of the fluid), and the logarithmic layer using the logarithmic law, i.e., 
\begin{equation}\label{eq:log-law}
    U^{+}=\frac{1}{\kappa}\ln\left( y^{+}\right) + 5.0, 
\end{equation}
where $\kappa = 0.4$ is the K\'{a}rm\'{a}n constant. 
A list of different piecewise formulae proposed in the literature for the law of the wall is shown in Table~\ref{tab:piecewise}. There are two disadvantages of the piecewise formulae: 1) the velocity in the buffer layer is not accurately described; 2) the velocity derivative is discontinuous because of the piecewise nature of these formulae. To resolve these two issues, different types of the law of the wall based on single formula have been proposed in the literature, which are often of two forms, i.e., the analytical form~\cite{Reichardt,Spalding,Rasmussen,Musker,Dean,Monkewitz} and the numerical form~\cite{Driest,Duprat}. The analytical forms are usually complicated or do not have a sound prediction~\cite{Spalding,Musker,Dean,Monkewitz}, while the numerical forms have a better prediction of the velocity profile~\cite{Driest,Duprat}, but do not have a clear physical meaning or difficult to use in practice. Different single formulae for the law of the wall proposed in the literature are shown in Table~\ref{tab:single1} and Table~\ref{tab:single2}.
%
\begin{table}
\centering
\begin{tabular}{ll}
\hline
\hline
\textbf{Authors} &  \textbf{Formulae}  \\
\hline
Prandtl~\cite{Prandtl} & \(  U^{+}=y^{+} \text{ for } 0 \leq y^{+} \leq 11.5 \)  \\
\hline
Taylor~\cite{Taylor} & \( U^{+}=2.5\ln  \left( y^{+} \right) +5.5 \text{ for } 11.5 \leq y^{+} \)    \\
\hline
\(\begin{matrix}
\\
\text{Von Karman}~\cite{Karman}\\
\\
\end{matrix}
\)
  & 
\(  \left\{ \begin{array}{l}
	U^{+}=y^{+}\text{, for } 0 \leq y^{+}<5\\
	U^{+}=5\ln  \left( y^{+} \right) -3.05\text{ for }  5 \leq y^{+}<30\\
	U^{+}=2.5\ln  \left( y^{+} \right) +5.5\text{ for }  30 \leq y^{+}
	\end{array} \right . \)  \\
\hline
\( 
\begin{matrix}
\\
\text{Deissler}~\cite{Deissler}\\
\\
\end{matrix}
 \)  & 
 \(  \left\{ \begin{array}{l}
	U^{+}= \int _{0}^{y^{+}} \left[ 1+n^{2}U^{+}y^{+} \left( 1-e^{-n^{2}U^{+}y^{+}} \right)  \right] ^{-1}dy^{+}\text{, where }n=0.124 \text{ for }0 \leq y^{+}<26 \\
U^{+}=2.78\ln  \left( y^{+} \right) +3.8\text{, for }26 \leq y^{+}\\
	\end{array} \right. \)  \\
\hline
 \( \begin{matrix}
\text{Rannie}~\cite{Rannie} \\
\end{matrix}
 \)  & 
\(  \left\{ \begin{array}{l}
	U^{+}=1.454\tanh  \left( 0.0688y^{+} \right) \text{ for } 0 \leq y^{+}<27.5\\
	U^{+}=2.5\ln  \left( y^{+} \right) +5.5 \text{ for } 27.5 \leq y^{+}\\
	\end{array}  \right. \)  \\
\hline
 \( \begin{matrix}
\\
\text{Breuer}\text{ \& }\text{Rodi} ~\cite{Breuer}\\
\\
\end{matrix}
 \)  & 
 \(  \left\{ \begin{array}{l}
	U^{+}=y^{+} \text{ for } 0 \leq y^{+}<5\\
	U^{+}=A\ln \left( y^{+} \right) +B \text{, } \\ \text{~~~~~~~~where }  A= \left[ k^{-1}\ln  \left( 30E \right) -5 \right] /\ln  \left( 6 \right)\text{, }B=5-A\ln \left( 5 \right) \text{ for } 5 \leq y^{+}<30\\
	U^{+}=k^{-1}\ln  \left( Ey^{+} \right) \text{, where }E=9.8 \text{ for } 30 \leq y^{+}
	\end{array}  \right. \)  \\
\hline
 \( \begin{matrix}
\text{Werner \& Wengle}~\cite{Werner}\\
\end{matrix}
 \)  & 
 \(  \left\{ \begin{array}{l}
	U^{+}=y^{+} \text{ for } 0 \leq y^{+}<11.81\\
	U^{+}=A \left( y^{+} \right) ^{B}\text{, where } A=8.3, B=1/7 \text{ for } 11.81 \leq y^{+}\\
	\end{array}  \right. \)  \\
\hline
\( \begin{matrix}
\\
\text{Inagaki et al.}~\cite{Inagaki}\\
\\
\end{matrix}
 \)  & 
 \(  \left\{ \begin{array}{l}
	U^{+}=y^{+} \text{ for } 0 \leq y^{+}<y_{C_1}^{+}\\
	U^{+}=A_1 \left( y^{+} \right) ^{B_1} \text{, where } A_1=2.7,B_1=1/2,y_{C_1}^{+}=A_1^{1/ \left( 1-B_1 \right) } \text{ for } y_{C_1}^{+} \leq y^{+}<y_{C_2}^{+}\\
	U^{+}=A_2 \left( y^{+} \right) ^{B_2} \text{, where } A_2=8.6,B_2=1/7,y_{C_2}^{+}= \left( A_2/A_1 \right) ^{1/ \left( B_1-B_2 \right) } \text{ for } y_{C_2}^{+} \leq y^{+}\\
	\end{array} \right. \)  \\
\hline
\hline

\end{tabular}
\caption{Formulae for the law of the wall: piecewise function.\label{tab:piecewise}}
 \end{table}
%

\begin{table}
\centering
\begin{tabular}{ll}
\hline
\hline
\textbf{Authors} & 
\textbf{Formulae} \\
\hline
Reichardt~\cite{Reichardt} & 
\( 
\begin{array}{l}
U^{+}=2.5\ln  \left( 1+0.4y^{+} \right) +7.8 \left( 1-e^{- \frac{y^{+}}{11}}-\frac{y^{+}}{11}e^{-0.33y^{+}} \right)  
\end{array}
\)  \\
\hline
Spalding~\cite{Spalding} & 
\( 
\begin{array}{l}
f \left( U^{+} \right) =U^{+}+e^{-A} \left( e^{\kappa U^{+}}-1-\kappa U^{+}-\frac{ \left( \kappa U^{+} \right) ^{2}}{2!}-\frac{ \left( \kappa U^{+} \right) ^{3}}{3!}-\frac{ \left( \kappa U^{+} \right) ^{4}}{4!} \right)    
\end{array}
\)
\\
\hline
Rasmussen~\cite{Rasmussen} & 
\( 
\begin{array}{l}
y^{+}=f \left( U^{+} \right) \text{, where } f \left( U^{+} \right) =U^{+}+e^{-A} \left( 2\cosh \left( \kappa U^{+} \right) - \left( \kappa U^{+} \right) ^{2}-2 \right),A=2.2, \kappa=0.4 
\end{array}
\) 
\\
\hline
Musker~\cite{Musker} & 
\( 
\begin{array}{l} 
U^{+}=5.424\tan ^{-1} \left[ \frac{2y^{+}-8.15}{16.7} \right] +\log _{10} \left[ \frac{ \left( y^{+}+10.6 \right) ^{9.6}}{ \left( {y^{+}}^{2}-8.15y^{+}+86 \right) ^{2}} \right] -3.52+ \\2.44 \left\{  \Pi  \left[ 6 \left( \frac{y}{ \delta } \right) ^{2}-4 \left( \frac{y}{ \delta } \right) ^{3} \right] + \left[  \left( \frac{y}{ \delta } \right) ^{2} \left( 1-\frac{y}{ \delta } \right)  \right]  \right\}\text{, }\Pi =0.55 
\end{array}
\)  \\
\hline
Dean~\cite{Dean} & 
\( 
\begin{array}{l} 
y^{+}e^{\kappa g \left(  \Pi ,~~ \frac{y}{ \delta } \right) }=f \left( U^{+} \right) ,g \left(  \Pi , \frac{y}{ \delta } \right) =\frac{1}{\kappa} \left( 1+6 \Pi  \right)  \left( \frac{y}{ \delta } \right) ^{2}-\frac{1}{\kappa} \left( 1+4 \Pi  \right)  \left( \frac{y}{ \delta } \right) ^{3} \\
\text{where } f \left( U^{+} \right) \text{is given by Spalding or Rasmussen's expressions listed above} \\
\end{array}
\)  \\
\hline
$\begin{array}{l}
    \text{Monkewitz,} \\
    \text{Chauhan} \\
    \text{and Nagib}~\cite{Monkewitz} 
\end{array}$
& 
\( 
\begin{array}{ll}
U_{\text{inner}}^{+}~~=&U_{\text{inner, 23}}^{+}+U_{\text{inner, 25}}^{+} \\
U_{\text{inner, 23}}^{+}= &
  0.68285472\ln \left( {y^{+}}^{2}+4.7673096y^{+}+9545.9963 \right) + \\ 
 & 1.2408249\arctan \left( 0.010238083y^{+}+0.024404056 \right) + \\ 
 & 1.2384572\ln \left( y^{+}+95.232690 \right) -11.930683  \\
 U_{inner,25}^{+}=&-0.50435126\ln \left( {y^{+}}^{2}-7.8796955y^{+}+78.389178 \right) + \\ 
 & 4.7413546\arctan \left( 0.12612158y^{+}-0.49689982 \right) \\ 
 & -2.7768771\ln \left( {y^{+}}^{2}+16.209175y^{+}+933.16587 \right) + \\ 
 & 0.37625729\arctan \left( 0.033952353y^{+}+0.27516982 \right) + \\ 
 & 6.5624567\ln \left( y^{+}+13.670520 \right) +6.1128254 
\end{array}
\)  \\
\hline
van Driest~\cite{Driest} & 
\( U^{+}= \int _{0}^{y^{+}}2 \left[ 1+\sqrt[]{1+0.64{y^{+}}^{2} \left( 1-e^{- \frac{y^{+}}{26}} \right) ^{2}} \right] ^{-1}dy^{+} \)  \\
\hline
Duprat et al.~\cite{Duprat} & 
\( 
\begin{array}{ll}
\frac{ \partial U^{+}}{ \partial y^{+}}=\frac{\text{sign} \left( \frac{ \partial P}{ \partial x} \right)  \left( 1- \alpha  \right) ^{3/2}y^{+}+\text{sign} \left(  \tau_{w} \right)  \alpha }{1+\frac{ \nu _{t}}{ \nu }}   \\
 \frac{ \nu _{t}}{ \nu }=ky^{+} \left[  \alpha +y^{+} \left( 1- \alpha  \right) ^{3/2} \right] ^{ \beta } \left( 1-e^{-y^{+}/ \left( 1+A \alpha ^{3} \right) } \right) ^{2}  \text{, where } \beta =0.78,A=17, \alpha =1   \\
\end{array}
\)\\
\hline
Cantwell~\cite{Cantwell2019}&
\(\begin{array}{ll}
    U^+(y^+)=\int_{0}^{y^+}\left(-\frac{1}{2\lambda(s)^2}+\frac{1}{2\lambda(s)^2}\left(1+4\lambda(s)^2\left(1-\frac{s}{R_\tau}\right)\right)^{1/2}\right)ds \\
    \lambda(y^+)=\frac{{\kappa}y^+\left(1-e^{-(y^+/a)^m}\right)}{\left( 1+\left( \frac{y^+}{bR_\tau}\right)^n\right)^{1/n}}\text{, where }k=0.4092,a=20.095,m=1.621,b=0.3195,n=1.619.
\end{array}\)\\
\hline
Rotta~\cite{Rotta1950Das}&
\(\begin{array}{ll}
    U^+=\frac{1}{2{\kappa}I^+_m}\left(1-\sqrt{1+4{I^+_m}^2} \right)+\frac{1}{\kappa}ln\left(2{I^+_m}+\sqrt{1+4{I^+_m}^2}\right)+{\delta}^+_l,\\
    \text{where }I_m={\kappa}(y-{\delta}_l)\text{, } I^+_m=u_{\tau}I_m/\nu \text{ and }{\delta}^+_l=u_{\tau}\delta_l/\nu\\
    {\delta}^+_l=5.0\text{ and experimental data~\cite{anderson1972turbulent,purtell1981turbulent} suggests }{\delta}^+_l=7.0.
\end{array}\)\\
\hline

\hline
\hline

\end{tabular}
\caption{Formulae for the law of the wall: single formula.\label{tab:single1}}
\end{table}
\begin{table}
\centering
\begin{tabular}{ll}
\hline
\hline
\textbf{Authors} & 
\textbf{Formulae} \\
\hline
Nickels~\cite{NICKELS2004Inner}&
\(\begin{array}{ll}
    U^+={y^+_c}\left[1-\left(1+2(y^+/y^+_c)+\frac{1}{2}({3-p^+_x}{y^+_c})(y^+/y^+_c)^2 -\frac{3}{2}{p^+_x}{y^+_C}(y^+/y^+_c)^3\right)e^{-3y^+/y^+_c}\right]+\\
    \frac{\sqrt{1+{p^+_x}{y^+_c}}}{6{\kappa}_o}\ln\left(\frac{1+(0.6(y^+/y^+_c))^6}{1+{\eta}^6}\right)+b\left(1-e^{-\frac{5({\eta}^4+{\eta}^8)}{1+5{\eta}^3}}\right),\text{where }{\eta}=y/\delta,p^+_x=(\nu/{\rho}U_\tau^3)/(dp/dx),\\
    p^+_x{y^+_c}^3+{y^+_c}^2-R^2_c=0,R_c=\frac{{U_T}{y_c}}{\nu}\text{ and } U_T=\sqrt{\tau(y=y_c)/\rho}
\end{array}\)\\
\hline
Haritonidis~\cite{Haritonidis1989A}&
\(\begin{array}{ll}
    U^+=\frac{1}{\lambda}\arctan{{\lambda}y^+}-\frac{a}{2{\lambda}^2}\ln\left({1+{\lambda}^2{y^+}^2}\right)\text{, where }{\lambda}^2={\alpha}f^+,\alpha=\frac{nm^2}{2},a=1/h^+,\\
    \text{n is the number of ejections of equal strength, }f=1/{\Delta}t_b\text{ is the bursting frequency,}\\
    m={\kappa}n^{-1}({\Delta}t_e/{\delta}t_b)^{-1}\text{ and ${\Delta}t_e$ is the duration of the ejections.}
\end{array}\)\\
\hline
Yakhot et al.~\cite{Yakhot1993Analytic}&
\(\begin{array}{ll}
    U^+(y^+)=\frac{1}{3\kappa}\left[4C^{1/4}-z+\ln\left(\frac{z+1}{z-1}\right)+2\arctan{(z)}-\ln{\left(\frac{C^{1/4}+1}{C^{1/4}-1}\right)}-2\arctan{(C^{1/4})}\right],\\
    \text{where }z=({\hat{\nu}}^3-1+C)^{1/4}/\hat{\nu}^{3/4}\text{  and }\hat{\nu}^4+(C-1)\hat{\nu}-\hat{\nu}^4_m=0,\\
    \hat{\nu}=\nu_t/\nu,\hat{\nu}_m={\kappa}l^+,l^+=u_{\tau}l/\nu.
\end{array}\)\\
\hline
Nikuradse~\cite{nikuradse1966laws}&
\(\begin{array}{ll}
    U^+=\int^{y^+}_{0} {2\left(1-\frac{y^+}{Re_{\tau}}\right)}\left[1+\sqrt{1+4{l_m^+}^2\left(1-\frac{y^+}{Re_{\tau}}\right)}\right]^{-1}dy^+,\\
    \frac{l_m}{\delta}=0.14-0.08\left(1-\frac{y^+}{Re_\tau}\right)^2-0.06\left(1-\frac{y^+}{Re_\tau}\right)^4
\end{array}\)\\
\hline
$\begin{array}{l}
    \text{Cebeci,} \\
    \text{Bradshaw}~\cite{Cebeci1984Physical}
\end{array}$&
\(\begin{array}{ll}
    U^+=\int^{y^+}_{0} {2\left(1-\frac{y^+}{R_{\tau}}\right)}\left[1+\sqrt{1+4{l_m^+}^2\left(1-\frac{y^+}{R_{\tau}}\right)}\right]^{-1}dy^+,\\
    \frac{l_m}{\delta}=\left(0.14-0.08\left(1-\frac{y^+}{Re_\tau}\right)^2-0.06\left(1-\frac{y^+}{Re_\tau}\right)^4\right)\left(1-e^{-\frac{y^+}{26}}\right)
\end{array}\)\\

\hline
\hline

\end{tabular}
\caption{Formulae for the law of the wall: single formula.\label{tab:single2}}
\end{table}
%
In this work, we propose a new single formula named as the logarithmic-exponential (LOG-EXP) formula shown as follows:
\begin{equation}\label{eq:the formula}
    U^{+} \left( y^{+} \right) =\frac{1}{\kappa}\ln  \left( 1+\kappa y^{+} \right) +A \left( 1-e^{-\frac{y^{+}}{B}} \right) +C \left( 1-e^{-\frac{y^{+}}{D}} \right),
\end{equation}
where $A=11.630$, $B=7.194$, $C=-4.472$ and $D=2.766$, validate the proposed formula using direct numerical simulation (DNS) data~\cite{Lee}, the experimental data from other canonical wall bounded flows~\cite{laufer1953,charnay1972,lindgren1969} and the classic law of the wall (i.e., Eq.~(\ref{eq:linear}) and Eq.~(\ref{eq:log-law})), and apply the new single formula to wall-modelled large-eddy simulation (WMLES) via a feedforward neural network (FNN) model for explicitly computing the wall shear stress using the wall-normal distance and streamwise velocity.

The rest of this paper is organized as follows: in section~\ref{sec:Derivation}, the derivation process of the LOG-EXP formula is presented; the proposed formula is validated using the DNS data of turbulent channel flows and the experimental data of other canonical wall bounded flows in section~\ref{sec:Validation}; then it is applied WMLES via a feedforward neural network in section~\ref{sec:FNN}; at last, conclusions are drawn in section~\ref{sec:Conclusions}.
%
%
\section{Derivation of the LOG-EXP formula of the wall}\label{sec:Derivation}
To derive the single formula for the law of the wall, we propose the following expression for the first derivative of the mean streamwise velocity,
\begin{equation}\label{eq:dUdy}
	\frac{dU^{+}}{dy^{+}}=\frac{1}{1+\kappa y^{+}}+ \Phi \left( y^{+} \right)
\end{equation}
where $\Phi\left( y^{+} \right)$ is the function to be given and satisfies these conditions, i.e., $\mathop{\lim }_{\mathop{y}^{+} \rightarrow \infty} \Phi \left( y^{+} \right) =\mathop{\lim }_{\mathop{y}^{+} \rightarrow 0} \Phi \left( y^{+} \right) =0$. It is noticed that this expression exactly satisfies the boundary condition at the wall, i.e., 
\begin{equation}
	\frac{dU^{+}}{dy^{+}}=1 \text{ at } y^{+} = 0,
\end{equation}
and approximately satisfies the condition in the logarithmic layer, i.e., 
\begin{equation}
	\frac{dU^{+}}{dy^{+}}=\frac{1}{\kappa y^{+}} \text{ for } y^{+} \text{in the logarithmic region,}
\end{equation}
where $y^{+}$ is large enough such that the first term on the right-hand side of Eq.~(\ref{eq:dUdy}) can be approximated by ${1}/\left({\kappa y^+}\right)$, and the second term on the right-hand side of Eq.~(\ref{eq:dUdy}) is considered being negligible.

There are many choices for  \(  \Phi \left( y^{+} \right)  \). In this work, we propose to use the following expression: 
\begin{equation}\label{eq:Phi}
	 \Phi \left( y^{+} \right) =\frac{A}{B}e^{\frac{-y^{+}}{B}}+\frac{C}{D}e^{\frac{-y^{+}}{D}}, 
\end{equation}
where $A$, $B$, $C$, and $D$ are constants satisfying the following constraint 
\begin{equation}\label{eq:constr}
    \frac{A}{B}+\frac{C}{D}=0.
\end{equation}
We will try to explain the physical meaning of the two exponential terms in Eq.~(\ref{eq:Phi}) after we set the values of $A$, $B$, $C$ and $D$. To obtain the law of the wall, we first substitute Eq.~(\ref{eq:Phi}) into Eq.~(\ref{eq:dUdy}), which yields
\begin{equation}\label{eq:dUdy2}
	\frac{dU^{+}}{dy^{+}}=\frac{1}{1+\kappa y^{+}}+\frac{A}{B}e^{\frac{-y^{+}}{B}}+\frac{C}{D}e^{\frac{-y^{+}}{D}}. 
\end{equation}
Then, the expression for $U^+$ is obtained by integrating Eq.~(\ref{eq:dUdy2}) with respect to \( y^{+} \) as follows:
\begin{equation}\label{eq:U}
	U^{+} \left( y^{+} \right) =\frac{1}{\kappa}\ln  \left( 1+\kappa y^{+} \right) +A \left( 1-e^{-\frac{y^{+}}{B}} \right) +C \left( 1-e^{-\frac{y^{+}}{D}} \right). 
\end{equation}
Finally, the values of $A$, $B$, $C$ and $D$ are determined by fitting Eq.~(\ref{eq:U}) using the DNS data~\cite{Lee} of the turbulent channel flow at $Re_{\tau}=5200$ with the constraint shown in Eq.~(\ref{eq:constr}). Specifically, the mean velocity values in the range of $0 \leq y \leq 0.1\delta$ are employed to obtain the parameters as follows: $A=11.630$, $B=7.194$, $C=-4.472$ and $D=2.766$. 

\begin{figure}
	\centering
		\includegraphics[width=0.65\textwidth]{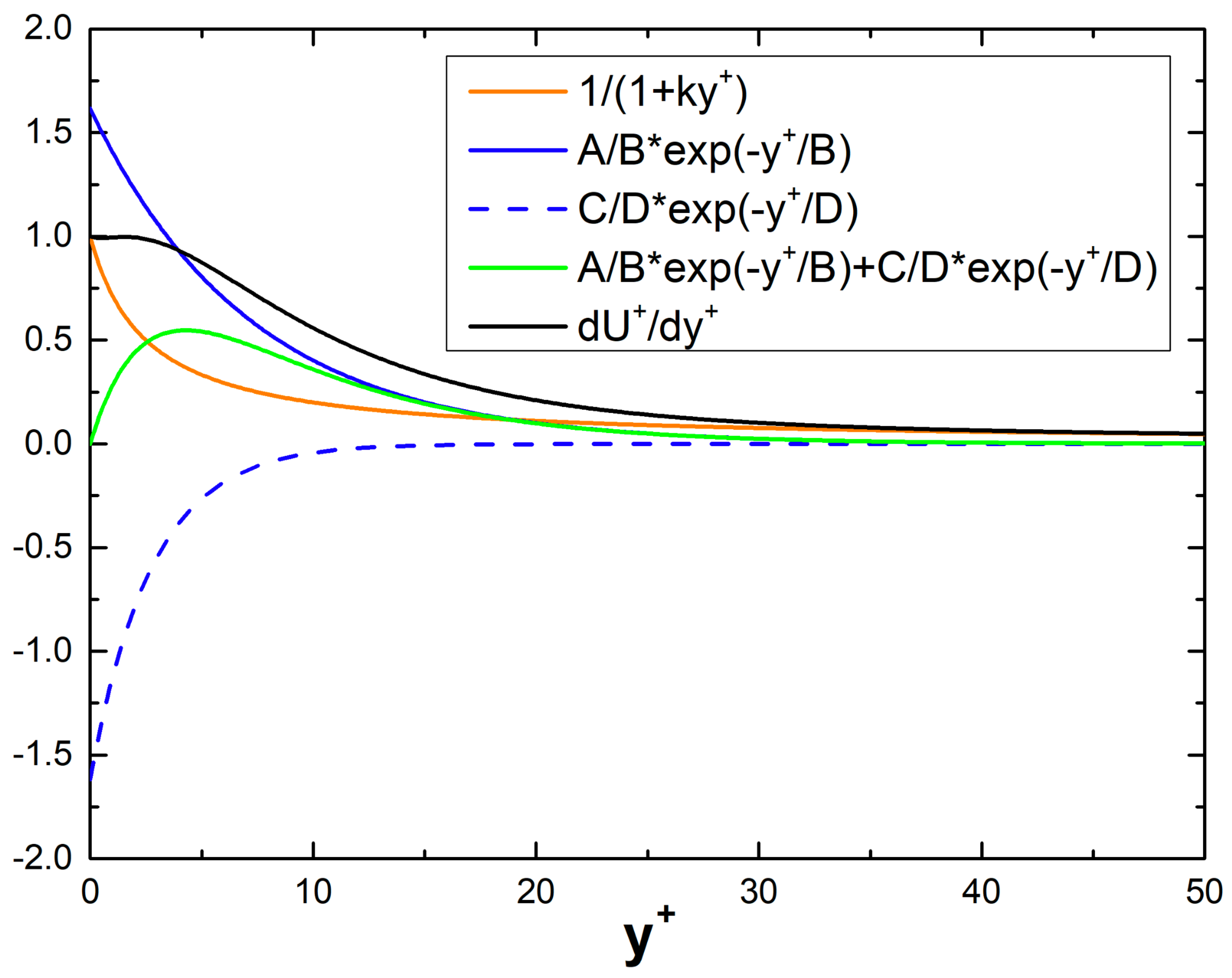}
	\caption{Vertical variations of different terms in Eq.~(\ref{eq:U}).}
	\label{fig:dUdy}
\end{figure}

We examine different terms in Eq.~(\ref{eq:dUdy2}) in figure~\ref{fig:dUdy}. As seen, the velocity derivative captured by the first term on the right-hand side of Eq.~(\ref{eq:dUdy2}), i.e., ${1}/{\left(1+\kappa y^{+}\right)}$, is less than the actual gradient. The second term on the right-hand side of Eq.~(\ref{eq:dUdy2}), i.e., the exponential term with a positive coefficient $\left({A}/{B}\right)e^{{-y^{+}}/{B}}$, which increases as approaching the wall, takes into account the effect of increased turbulence as approaching the buffer layer region. The third term on the right-hand side of Eq.~(\ref{eq:dUdy2}), i.e., the exponential term with a negative coefficient $\left({C}/{D}\right)e^{{-y^{+}}/{D}}$, on the other hand, decreases as approaching the wall, which offsets the increase of the second term, and acts as a damping function in the near-wall region. It is noticed that we attempt to understand the effects of different terms in Eq.~(\ref{eq:dUdy2}) instead of the exact physical interpretation of different terms. 

The asymptotic behavior of the derived single formula is examined as follows. In the viscous sublayer, we expand  Eq.~(\ref{eq:U}) in Taylor series about  $y^{+}=0$ and neglect the high order terms for small \( y^{+} \)  and obtain the linear profile as follows: 
\begin{equation}
	\mathop{\lim }_{\mathop{y}^{+} \rightarrow 0}U^{+} \left( y^{+} \right) \approx \mathop{\lim }_{\mathop{y}^{+} \rightarrow 0} \left( y^{+}+\frac{A}{B}y^{+}+\frac{C}{D}y^{+} \right) =y^{+}.
\end{equation}
In the logarithmic region, where \( y^{+} \)  is large enough at high Reynolds numbers, the logarithmic law can be recovered from Eq.~(\ref{eq:U}) shown as follows: 
\begin{equation}
	\mathop{\lim }_{\mathop{y}^{+} \rightarrow \infty}U^{+} \left( y^{+} \right) \approx \frac{1}{\kappa} \cdot \ln  \left( y^{+} \right) +A-\frac{D \cdot A}{B}+\frac{\ln  \left( \kappa \right) }{\kappa}=\frac{1}{\kappa}\ln \left( y^{+} \right) +4.9
\end{equation}
considering that  \( \mathop{\lim }_{\mathop{y}^{+} \rightarrow \infty}\ln  \left( \kappa+y^{+} \right) \approx \ln  \left( y^{+} \right)\) and \(\mathop{\lim }_{\mathop{y}^{+} \rightarrow \infty}e^{-\frac{y^{+}}{B}} \approx \mathop{\lim }_{\mathop{y}^{+} \rightarrow \infty}e^{-\frac{y^{+}}{D}} \approx 0 \). 
%
%
\section{Validation of the LOG-EXP formula}\label{sec:Validation}
In this section we validate the proposed LOG-EXP formula, i.e., Eq.~(\ref{eq:U}), using the DNS data of turbulent channel flow~\cite{Lee}, the experimental data from other canonical wall bounded flows~\cite{laufer1953,charnay1972,lindgren1969}.

\begin{figure}
	\includegraphics[width=\textwidth]{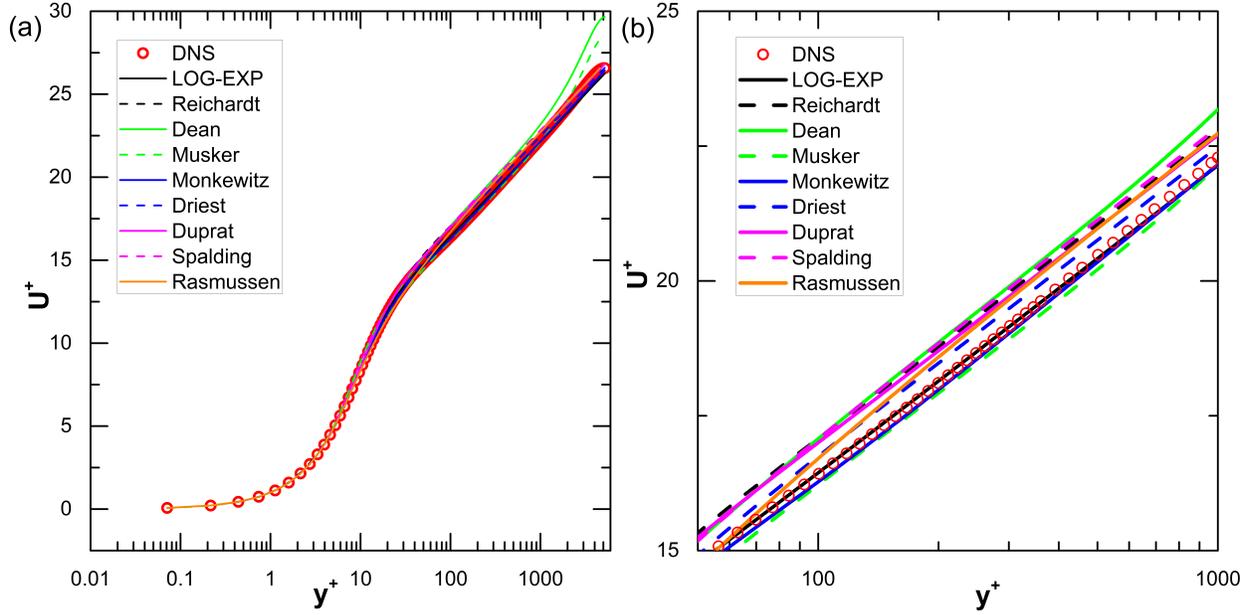}
	\caption{(a) Comparison of the velocity profiles predicted by the LOG-EXP formula and others with that from DNS for the turbulent channel flow at Reynolds numbers  \( Re_{ \tau}=5200 \) (b) Zoomed-in plot of (a) for $50 \leq y^{+} \leq 1000$.}
	\label{fig:U5200}
\end{figure}

\begin{figure}
	\includegraphics[width=0.75\textwidth]{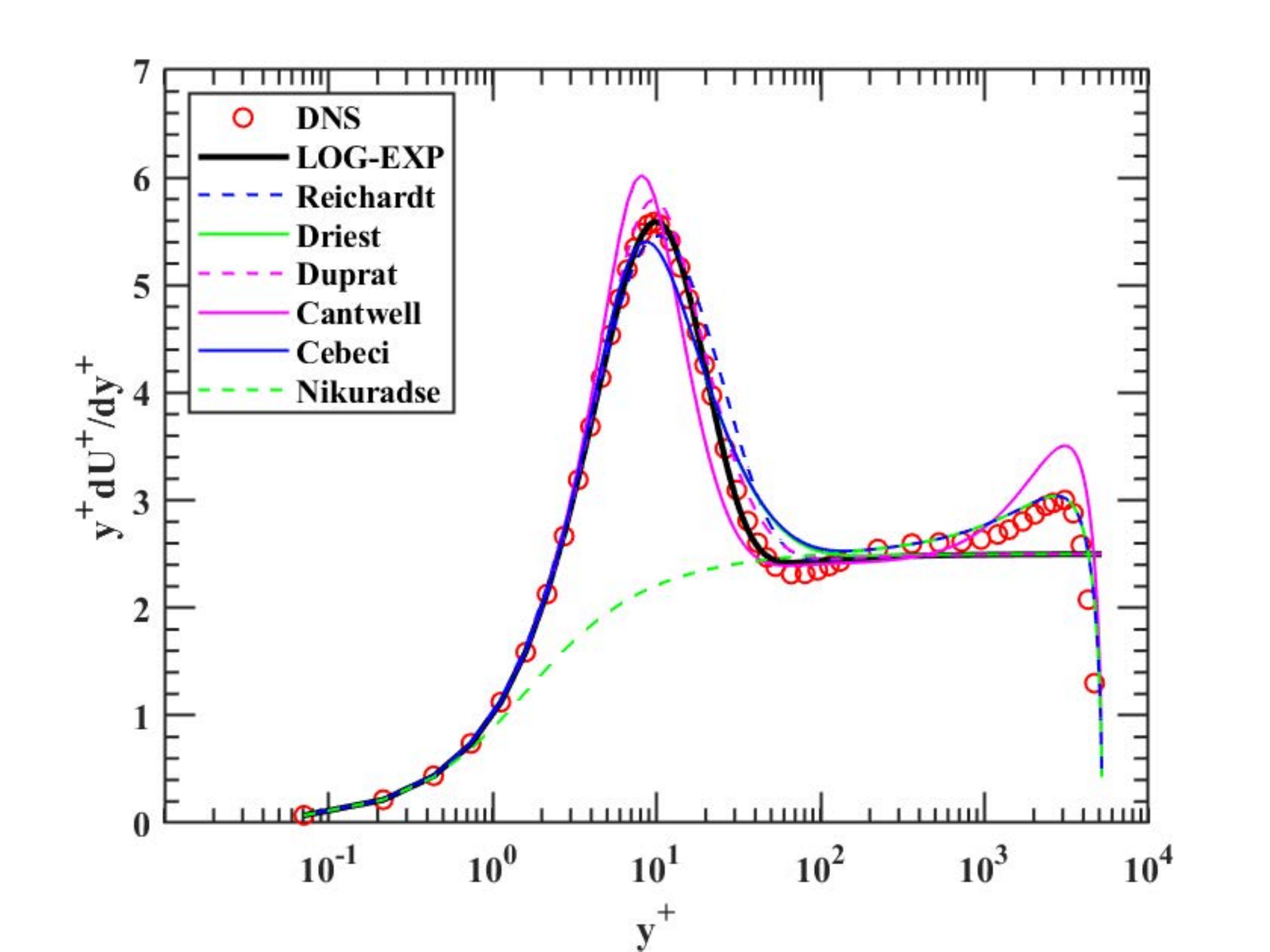}
	\caption{Comparison of $y^+dU^+/dy^+$ predicted by the LOG-EXP formula and others with that from DNS for the turbulent channel flow at Reynolds numbers  \( Re_{ \tau}=5200 \).}
	\label{fig:ydUdy5200}
\end{figure}

Figure~\ref{fig:U5200}(a) compares the predictions from several single formulae with the DNS data at \( ~Re_{ \tau}=5200 \), which are employed for calibrating the proposed formula. It is shown that all the formulae have a good prediction in the viscous sublayer and buffer layer. In the logarithmic region, which is enlarged and shown in Figure~\ref{fig:U5200}(b), the predictions of the proposed LOG-EXP formula collapse well with the DNS data, while some formulae, such as those predicted by Dean~\cite{Dean}, Reichardt~\cite{Reichardt}, Spalding~\cite{Spalding} and van Driest~\cite{Driest}, overpredict the velocity in this region. Furthermore, it is observed the predictions by Dean~\cite{Dean} and Musker~\cite{Musker} tend to deviate away from the DNS data. In Figure~\ref{fig:ydUdy5200}, we show the comparison of the first derivative of the velocity. As seen, predictions from different formulae agree well with the DNS data in the logarithmic layer and the outer layer. Discrepancies are observed in the viscous sublayer and buffer layer (i.e., $y^{+}<10$) for Reichardt’s formula~\cite{Reichardt} and Dean’s formula~\cite{Dean}, which overpredict the velocity derivative  \( dU^{+}/dy^{+} \). The velocity derivative predicted by the proposed LOG-EXP formula, on the other hand, agrees well with DNS data for all vertical locations. 

\begin{figure}
	\includegraphics[width=\textwidth]{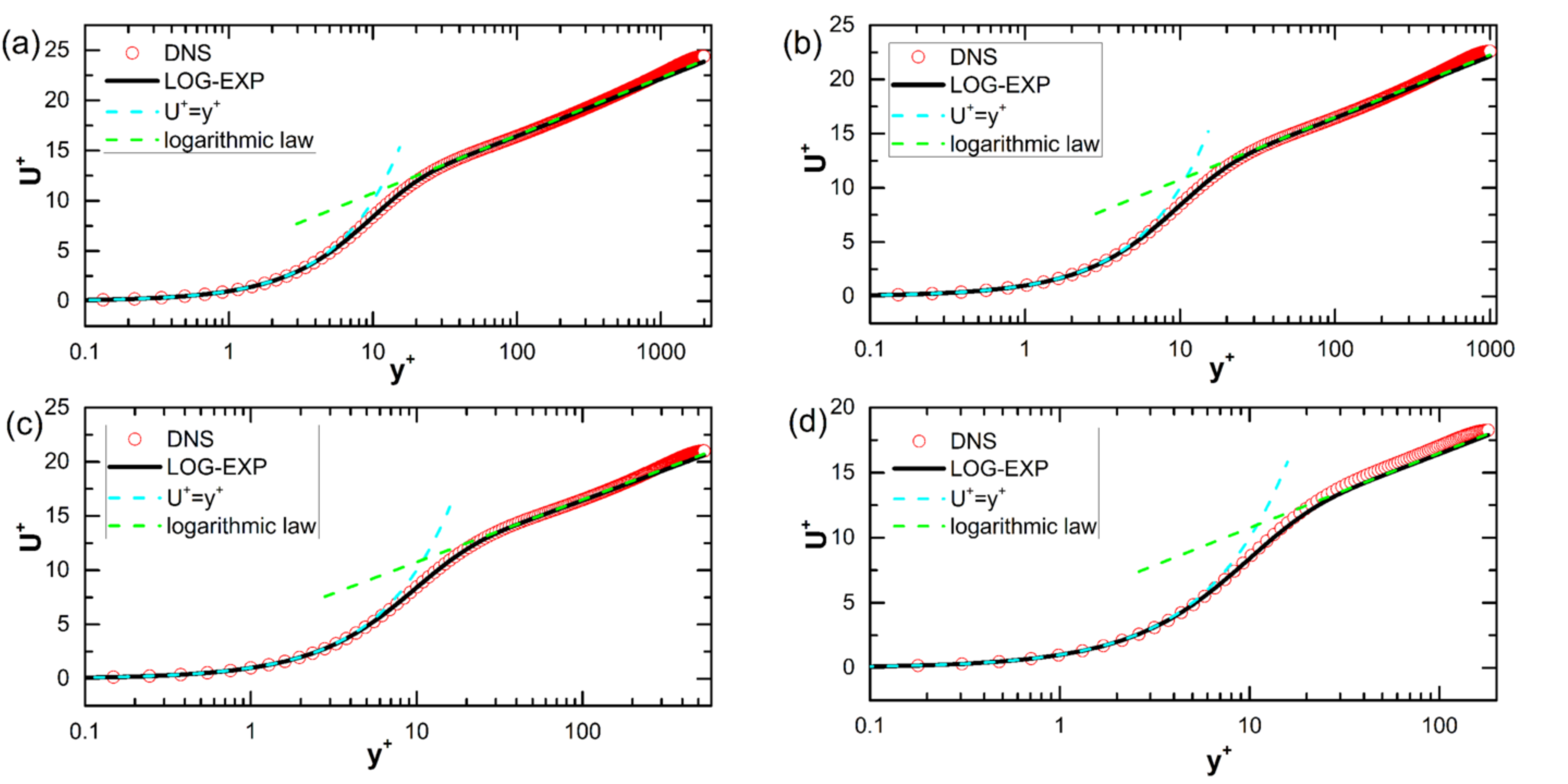}
	\caption{Comparison of the velocity profiles predicted by the LOG-EXP formula with those from DNS for the turbulent channel flow with different Reynolds numbers for (a) \( Re_{ \tau}=2000 \), (b) \( Re_{ \tau}=1000 \), (c) \( Re_{ \tau}=550 \), and (d) \( Re_{ \tau}=180 \).}
	\label{fig:UdiffRe}
\end{figure}

\begin{figure}
	\includegraphics[width=\textwidth]{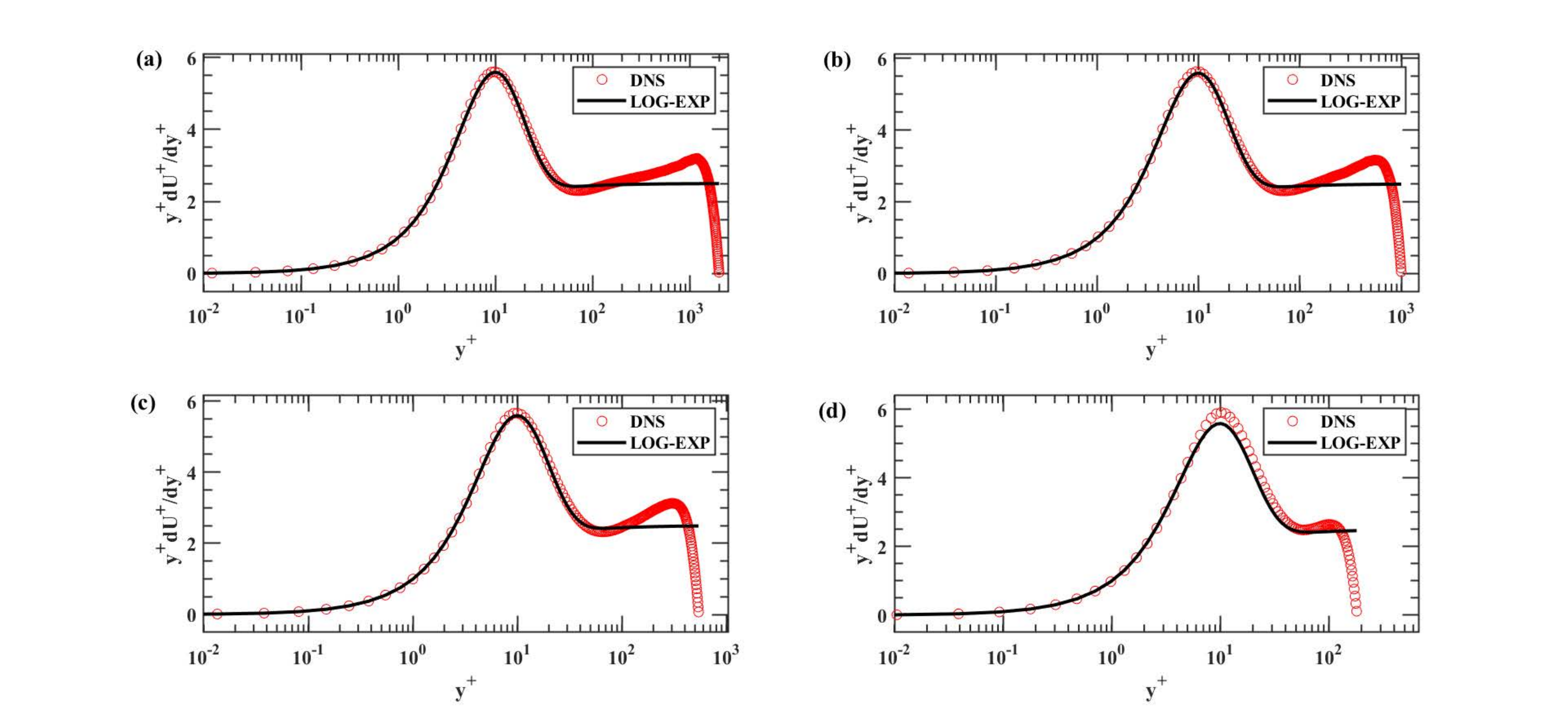}
	\caption{Comparison of $y^+dU^+/dy^+$ predicted by the LOG-EXP formula with those from DNS for the turbulent channel flow with different Reynolds numbers for (a) \( Re_{ \tau}=2000 \), (b) \( Re_{ \tau}=1000 \), (c) \( Re_{ \tau}=550 \), and (d) \( Re_{ \tau}=180 \).}
	\label{fig:ydUdydiffRe}
\end{figure}

In Figure~\ref{fig:UdiffRe} and Figure~\ref{fig:ydUdydiffRe}, we compare predictions from the proposed LOG-EXP formula with the DNS data~\cite{Lee} for different Reynolds numbers. Although minor differences are observed for the low Reynolds number case with $Re_{\tau}=180$, an overall good agreement is observed for both the velocity profile and the first derivative of $dU^{+}/dy^{+}$ for the proposed LOG-EXP formula.
%

To qualitatively evaluate the proposed LOG-EXP formula, we calculate the maximum absolute error of the velocity, which is defined by
\begin{equation}
  	e_{\text{max, }U^{+}}=\mathop{\max }_{0\leq\frac{y}{ \delta }\leq0.3}\frac{ \vert U_{F}^{+}-U_{DNS}^{+} \vert }{ \langle \left\vert U_{DNS}^{+} \right\vert \rangle }, 
\end{equation}
where  \( U_{DNS}^{+} \) , \( ~U_{F}^{+} \)  and  \(  \langle U_{DNS}^{+} \rangle  \)  are the velocity of the DNS data, the velocity calculated by the single formula and the mean of  \( U_{DNS}^{+} \)  averaged over  \( 0.0 \leq {y}/{ \delta } \leq 0.3 \), and the maximum absolute error of  \( dU^{+}/dy^{+} \) , which is defined by
\begin{equation}
  	e_{\text{max,}dU^+/dy^+}=\mathop{\max}_{0\leq\frac{y}{\delta}\leq0.3}\frac{ \left\vert  \left( \frac{dU^{+}}{dy^{+}} \right) _{F}—\left( \frac{dU^+}{dy^+} \right) _{DNS} \right\vert }{ \left\langle  \left\vert \left( \frac{dU^{+}}{dy^{+}} \right) _{DNS} \right\vert \right \rangle }, 
\end{equation}
where  \(  \left( dU^{+}/dy^{+} \right) _{DNS} \) ,  \(  \left( dU^{+}/dy^{+} \right) _{F} \)  and  \(  \langle  \left( dU^{+}/dy^{+} \right) _{DNS} \rangle  \)  are the first derivative of the velocity of the DNS data, the first derivative of the velocity calculated from single formula and the mean value of  \(  \left( dU^{+}/dy^{+} \right) _{DNS} \)  averaged over  \( 0.0 \leq {y}/{ \delta } \leq 0.3 \), respectively. 

\begin{figure}
	\includegraphics[width=\textwidth]{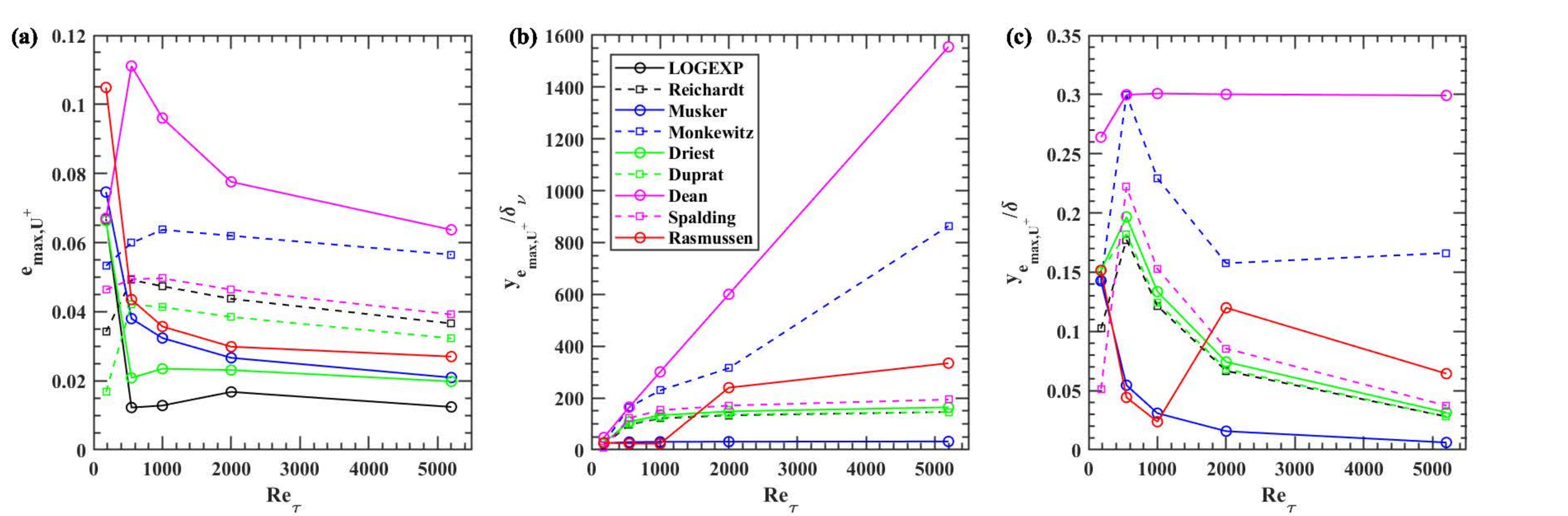}
	\caption{(a) The maximum absolute error $e_{\text{max, }U^{+}}$ of the velocity predicted by different formulae compared with DNS data for different Reynolds numbers; (b) The vertical location in wall units $y_{e_{\text{max, }U^{+}}}/\delta_v$ where the maximum absolute error occurs; (c) The vertical location normalized by the half-height of channel $y_{e_{\text{max, }U^{+}}}/\delta$ where the maximum absolute error occurs. }
	\label{fig:error_U}
\end{figure}

In Figure~\ref{fig:error_U}, we plot the errors of the velocity predicted by different formulae for different Reynolds numbers. As seen in Figure~\ref{fig:error_U}(a), the errors of the LOG-EXP predictions, which are less than 0.1, are lower than the other formulae except for the very low Reynolds number case. Figure~\ref{fig:error_U}(b) and (c) show the vertical location $y_{e_{\text{max, }U^{+}}}$ where the $e_{\text{max, }U^{+}}$ occurs for different normalization, i.e., the viscous length scale $\delta_v$ and the channel half-height $\delta$, respectively. It is observed that $y_{e_{\text{max, }U^{+}}}$ increases with the increase of the Reynolds number when normalized using $\delta_v$, while remains nearly constant for $y_{e_{\text{max, }U^{+}}}$ when normalized using $\delta$ for the proposed LOG-EXP formula,

\begin{figure}
	\includegraphics[width=\textwidth]{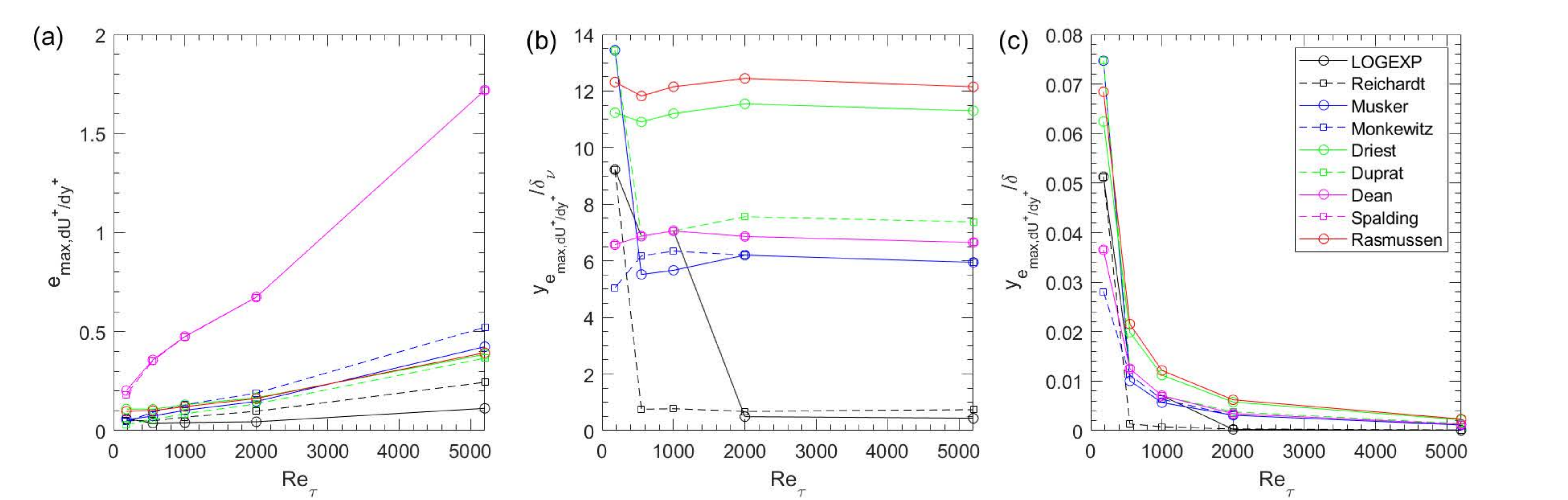}
	\caption{Same as Figure~\ref{fig:error_U} but for $e_{\text{max, }dU^{+}/dy^{+}}$ and $y_{e_{\text{max, }dU^{+}dy^{+}}}$.}
	\label{fig:error_dUdf}
\end{figure}

In Figure~\ref{fig:error_dUdf}, the errors of the velocity derivatives predicted by different formulae are examined for different Reynolds numbers. As seen in Figure~\ref{fig:error_dUdf}(a), the errors of the LOG-EXP predictions are lower than others, which increase via Reynolds number. Figure~\ref{fig:error_dUdf} (b) and (c) show the locations where the $e_{\text{max, }dU^{+}/dy^{+}}$ occurs, which are normalized by the length scales $\delta_v$ and $\delta$, respectively. As seen, the $y_{e_{\text{max, }dU^{+}dy^{+}}}$ for velocity derivative is close to or within the viscous sublayer for different Reynolds numbers.  
%


%

\begin{figure}
	\includegraphics[width=\textwidth]{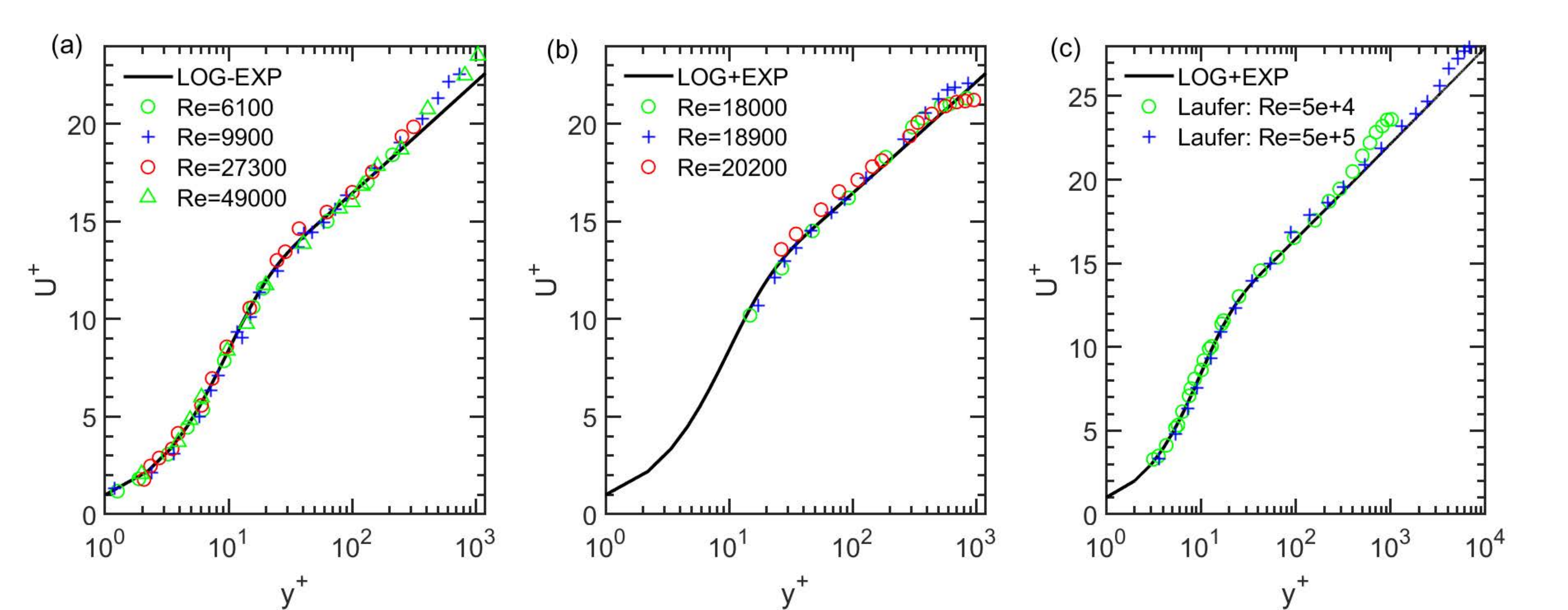}
	\caption{Comparison of the velocity profiles predicted by the LOG-EXP formula with the experimental data from (a) Lindgren and Chao~\cite{lindgren1969},(b) Charnay et al.~\cite{charnay1972}, (c) Laufer~\cite{laufer1953} for (a,c) the circular pipe flow and (b) the turbulent boundary layer on a flat plate. }
	\label{fig:Uothers}
\end{figure}

We further apply the proposed LOG-EXP formula to different types of canonical flows. Figure~\ref{fig:Uothers} shows the semi-logarithmic plots of mean velocity profiles in the circular pipe flow~\cite{laufer1953,lindgren1969} and the turbulent boundary layer on a flat plate~\cite{charnay1972}. An overall good agreement between the LOG-EXP formula and the experimental data is observed in the near wall region.

\section{Application of the LOG-EXP formula to WMLES}\label{sec:Application2WMLES}
In this section, we apply the LOG-EXP formula to WMLES. In WMLES, the wall shear stress at the wall is often employed as approximate boundary conditions for outer flow simulations. However, the wall shear stress cannot be determined explicitly using the LOG-EXP formula. To address this issue, we employ the feedforward neural network (FNN) to construct an explicit LOG-EXP model for explicitly computing the wall shear stress (details on the model training and a priori test can be found in the appendix~\ref{sec:FNN}. The explicit LOG-EXP model is then applied in WMLES and tested using turbulent channel flows at different Reynolds numbers.  

The explicit LOG-EXP model is implemented in the large-eddy simulation module of the Virtual Flow Simulator (VFS-Wind) code ~\cite{yang2015large,yang2018new}. The governing equations are the incompressible Navier-Stokes equations, which are discretized spatially using the second-order central differencing scheme, and advanced in time using a second-order  fractional step method. The subgrid-scale stress is modeled using the dynamic procedure~\cite{germano1991dynamic}. In the implementation of the LOG-EXP wall model, the wall-tangential component of the velocity and wall-normal distance as the second off-wall grid nodes are employed as inputs for the LOG-EXP model, with computed wall shear stresses employed as boundary conditions for the outer flow simulation. For the wall-normal component of the velocity, a no-slip boundary condition is employed. 

Two cases with different Reynolds numbers are considered, i.e., the Reynolds number defined using the friction velocity $Re_\tau=1000$, $5200$. For both cases, the computational domain is $L_x \times L_y \times L_z=7.0m \times 2.0m \times 3.5m$ in the streamwise, vertical and spanwise directions, respectively, with the corresponding numbers of grid nodes $N_x \times N_y \times N_z=33 \times 33 \times 33$, which are uniformly distributed in all three directions.  

%
%

\begin{figure}
	\includegraphics[width=\textwidth]{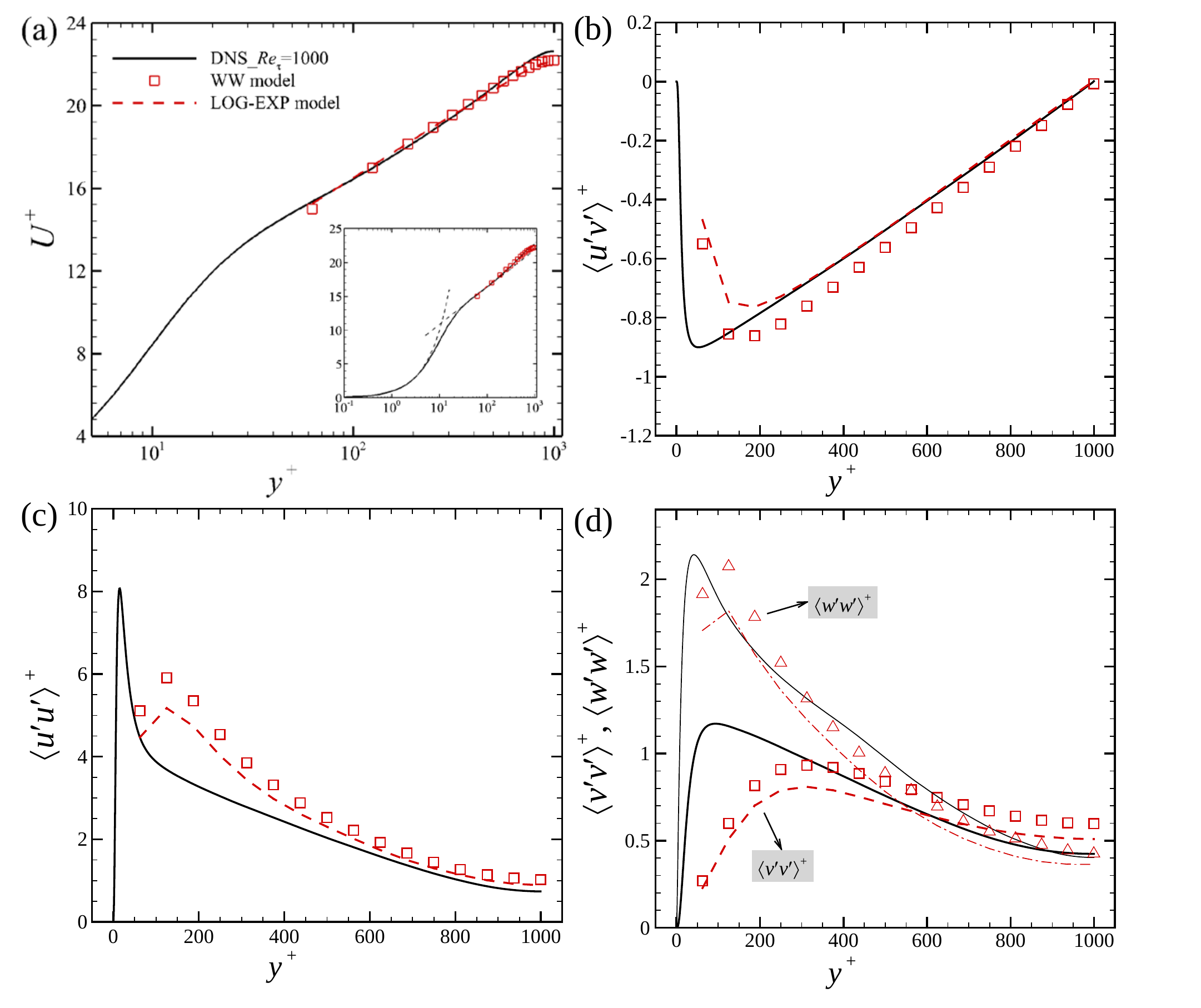}
	\caption{Comparison of predictions from the WMLES with the LOG-EXP model with DNS and those from the WW model for (a) the mean streamwise velocity, (b) the primary Reynolds shear stress ${\left< u'v'\right>}^+$, and (c, d) the normal Reynolds stresses ${\left< u'u'\right>}^+$, ${\left< v'v'\right>}^+$ and ${\left< w'w'\right>}^+$, respectively.}
	\label{fig:FNN_Retau1000}
\end{figure}

\begin{figure}
	\includegraphics[width=\textwidth]{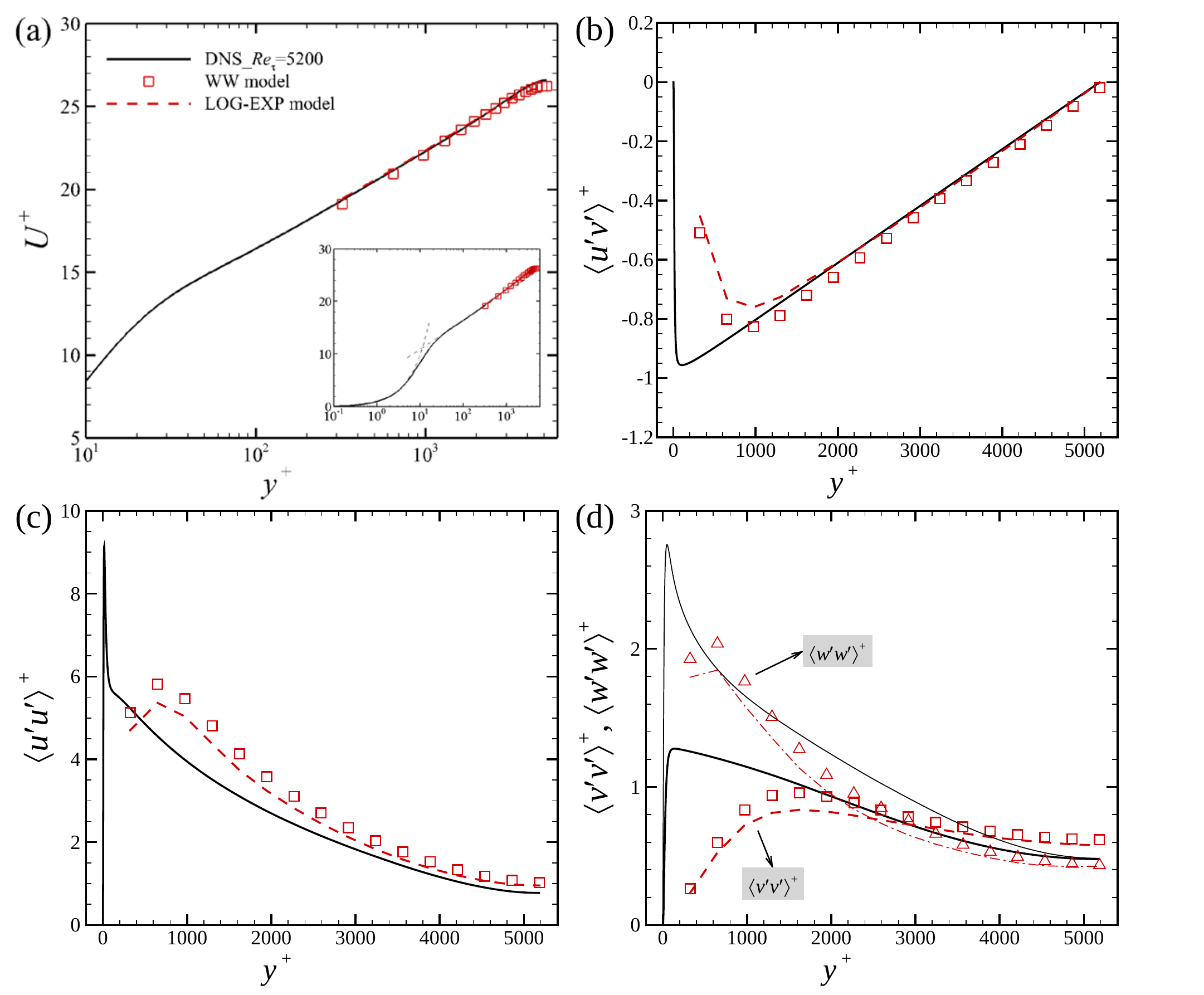}
	\caption{The same as figure ~\ref{fig:FNN_Retau1000} but for $Re_{\tau}=5200$.}
	\label{fig:FNN_Retau5200}
\end{figure}

We compare the predictions from WMLES with the LOG-EXP model with the DNS results~\cite{graham2016web} and those from the Werner-Wengle (WW) model~\cite{Werner}. It is shown in Figure~\ref{fig:FNN_Retau1000} (a) that the mean velocity profiles predicted by the WW model and the LOG-EXP model agree well with the DNS profile. For the primary Reynolds shear stress ${\left<u'v'\right>}^+$, figure~\ref{fig:FNN_Retau1000} (b) shows that the WW model somewhat overestimates the magnitude of ${\left<u'v'\right>}^+$ for $200<y^+<800$, which, on the other hand, is accurately predicted by the LOG-EXP model. The comparisons of the normal Reynolds stresses are shown in Figure~\ref{fig:FNN_Retau1000} (c, d). It is observed that both wall models overpredict the streamwise component of the normal Reynolds stress ${\left<u'u'\right>}^+$, while underpredict the wall-normal component ${\left<v'v'\right>}^+$. Comparisons for the case with $Re_{\tau}=5200$ are shown in Figure~\ref{fig:FNN_Retau5200}, with similar observations as for the case with $Re_{\tau}=1000$.
%
\section{Conclusions}\label{sec:Conclusions}
In this work, we proposed a new single formula for the law of the wall, which is dubbed as the LOG-EXP formula. It is derived based on the decomposition of the velocity derivative into a term satisfying the boundary conditions of the velocity derivative at the wall and in the logarithmic region, and two exponential terms accounting for the increase of turbulence as approaching the buffer layer and the viscous damping effect in the near-wall region, respectively. The derived LOG-EXP formula, which is composed of a logarithmic term and two exponential terms, works for the whole inner layer region and gives continuous predictions of both velocity and velocity derivative. Its capability is evaluated using DNS data of turbulent channel flows and some other canonical wall-bounded turbulent flows at different Reynolds numbers. Good agreements are obtained for all the considered cases. 

The proposed LOG-EXP formula is then applied to WMLES. To avoid the cost of solving the LOG-EXP expression for the friction velocity, an explicit LOG-EXP model is developed using the feedforward neural network for computing wall shear stress for WMLES. The WMLES with the LOG-EXP model is evaluated using the DNS data of turbulent channel flows. An overall good agreement is obtained for both mean streamwise velocity and the Reynolds stresses. 


%
\section*{Acknowledgements}
This work is partially supported by NSFC Basic Science Center Program for ``Multiscale Problems in Nonlinear Mechanics'' (NO. 11988102). 

\section*{Data availability}
The data that support the findings of this study are available from the corresponding author upon reasonable request.

\section*{Appendix A: A neural network model for explicitly computing wall shear stress}\label{sec:FNN}
\begin{figure}
    \centering
    \includegraphics[width=9cm]{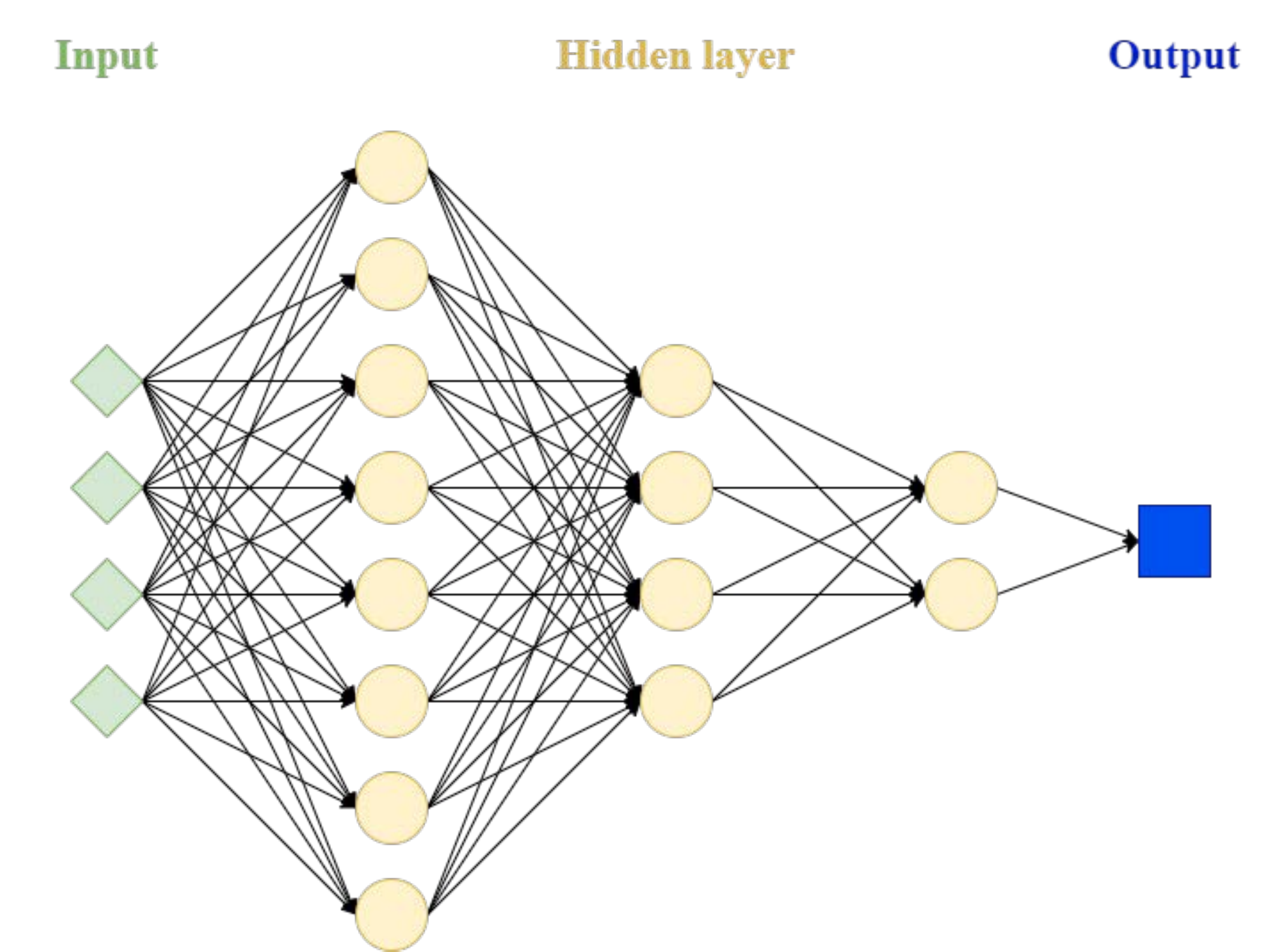}
    \caption{Schematic diagram of the feedforward neural network (FNN) with multi hidden layers.}
    \label{fig:NeuralNetwork}
\end{figure}

In this appendix, we describe the procedure for constructing a neural network model for explicitly computing the wall shear stress using the wall-normal distance and the streamwise velocity, which can be used directly in WMLES without the need to solve the implicit LOG-EXP formula. As shown in Figure~\ref{fig:NeuralNetwork}, a multi-hidden-layer feedforward neural network is employed for the construction of the explicit model, which contains an input layer, multi hidden layers and an output layer. Each layer has a number of neurons, which are computational units that take weighted sums of the inputs to an activation function and calculate the output. Then the outputs of each layer are fed forward
as inputs to the next layer. The training goal of the neural network is to find the optimal weight and bias coefficients to minimize the loss of the neural network. More details about the neural network can be found in the reference~\cite{goodfellow2016deep}.

The structure of the neural network employed in this work is shown in Tabel~\ref{tab:FNN}. The activation function used in this paper is the hyperbolic tangent function (tanh), which is defined as follows:
\begin{equation}
    f(x)=\frac{e^x-e^{-x}}{e^x+e^{-x}}.
\end{equation}
The input and output data are normalized using the Min-Max scaling,
\begin{equation}
    x^*=\frac{x-x_{min}}{x_{max}-x_{min}}.
\end{equation}
The loss function of the neural network is set as the mean square error (MSE), which is defined as follows:
\begin{equation}
    L_{FNN}=\frac{1}{N_s}\sum_{i=1}^{N_s}\left(y_i-y_i^*\right)^2,
\end{equation}
where $N_s$ is the number of training samples, $y_i$ and $y_i^*$ are the FNN output and the labeled output from the training data, respectively. Besides, the error back-propagating (BP) scheme~\cite{rumelhart1986learning} and the Adam optimizer~\cite{kingma2014adam} are implemented with Keras~\cite{chollet2018deep} to train the FNN model.
%

%
%

The training data is obtained by solving the Eq.~(\ref{eq:the formula}) about the friction velocity $u_{\tau}$ within the wall-normal distance {\color{black} in the range of $10^{2}<y/\nu<10^{5}$ and the streamwise velocity in the range of $0<U<1$. Data in $y^+< 10$ are precluded because no wall model would be needed when the small scales are also resolved. Data above $y/\delta=0.1$ are not included for training because the flow away from the wall is usually resolved by LES grids. Therefore, we have $10/u_\tau<y/\nu<0.1Re_\tau/u_\tau$. For comparison purposes, we assume $u_\tau{\approx}0.05$ and $0<U<1$, which are the same as the DNS data. Then we get  the approximate range of $y/\nu$. 501 points are selected evenly in the streamwise velocity range and 201 points are selected evenly in two adjacent orders of magnitude of $y/\nu$, so that the training data contains about 501 × 603 input-output pairs. The neural network is trained to predict the friction velocity for given velocity and wall-normal distance. The inputs for training the neural network, which are defined based on the expression of the LOG-EXP formula as shown in Eq.~\ref{eq:the formula}, are shown in Table ~\ref{tab:FNN}.

%
%
%

Figure~\ref{fig:FNN} (a) shows the friction velocity predicted by the FNN model, as a function of the streamwise velocity and the wall-normal distance divided by the kinematic viscosity. As seen in Figure~\ref{fig:FNN} (b), the relative error of the friction velocity, which is define by
\begin{equation}\label{eq:u_tau_RE}
  	{u_{\tau}}_{RE}=\frac{|u_{\tau}-{u_{\tau}}_{FNN}|}{u_{\tau}},
\end{equation}
where $u_{\tau}$ and ${u_{\tau}}_{FNN}$ are calculated by Eq.~\ref{eq:the formula} and FNN, respectively, is smaller than $0.5\%$ for most cases. And the relative error of the velocity in wall units obtained by FNN is smaller than $1\%$, except for the case when the streamwise velocity is very small. Then we test the FNN model using the DNS data of turbulent channel flow with  $Re_{\tau}=1000$~\cite{graham2016web}. As seen in Figure~\ref{fig:FNN_yplus=10_and_100}, the predictions from the FNN model fit well with the DNS data for different wall-normal distannces.
%
\begin{table}
    \centering
    \begin{tabular}{ccccc}
        \hline
        \hline
        \textbf{NN} & \textbf{ AF } & \textbf{HL size} & \textbf{Input} & \textbf{Output} \\
        \hline
        \textbf{FNN} & \textbf{ tanh } & ( 8 ,4 ,2 ) &\(\frac{1}{\kappa}ln(1+\frac{{\kappa}yU}{\nu}), A(1-e^{-\frac{yU}{{\nu}B}}), C(1-e^{-\frac{yU}{{\nu}D}}), \sqrt{\frac{yU}{\nu}} \) & \( \frac{U}{u_{\tau}} \) \\
        \hline
        \hline
    \end{tabular}
    \caption{Details of the neural networks. Here NN denotes neural network, AF activation function and HL hidden layer. The tabulated hidden layer size contains the number of neurons within each hidden layer.\label{tab:FNN}}
\end{table}
%
\begin{figure}
	\includegraphics[width=\textwidth]{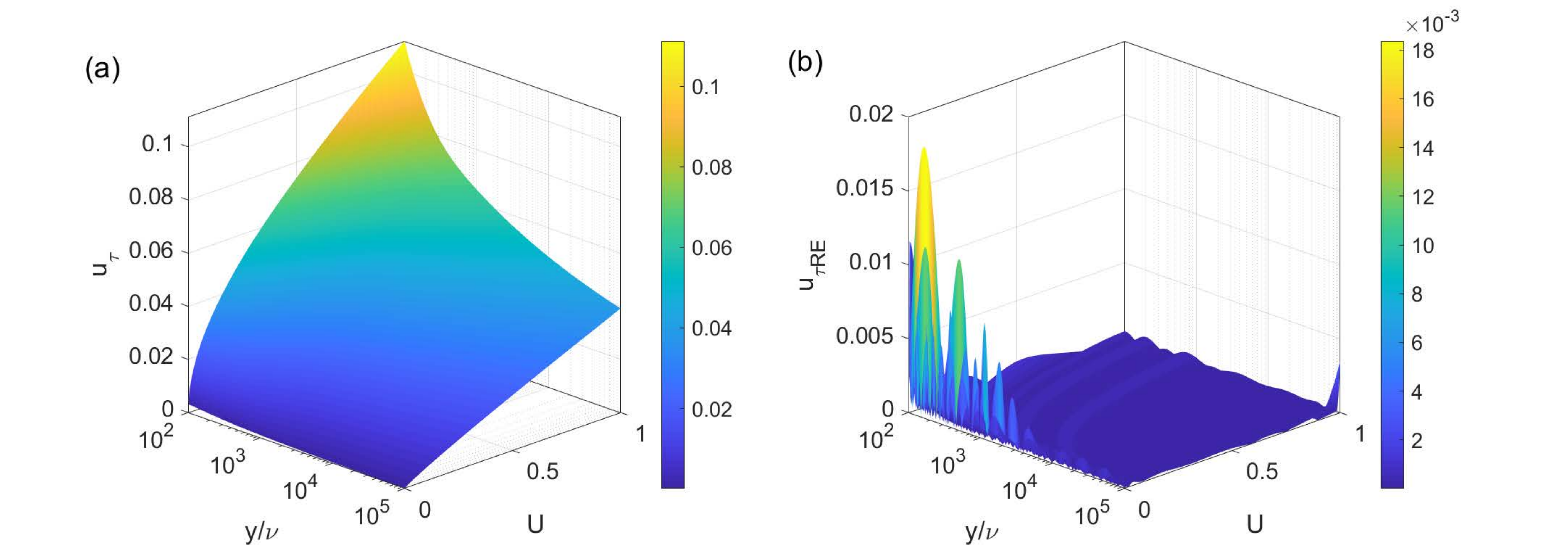}
	\caption{Test of the FNN model using the data from the LOG-EXP formula for (a) the friction velocity and (b) relative error of the friction velocity predicted by FNN.}
	\label{fig:FNN}
\end{figure}
\begin{figure}
	\includegraphics[width=12cm]{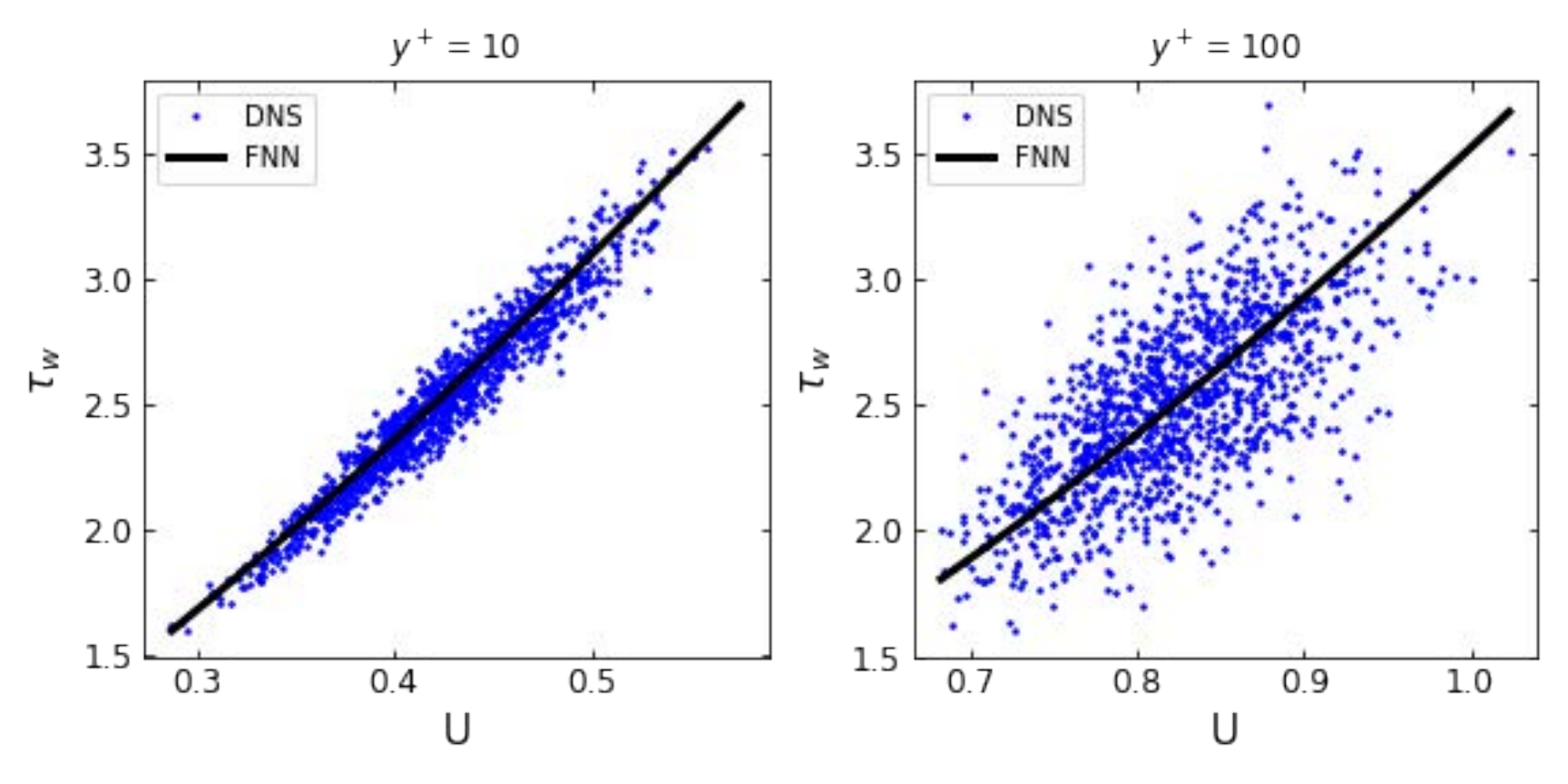}
	\caption{Test of the FNN model using the DNS data of turbulence channel flow at $Re_{\tau}=1000$ for (a) Wall shear stress as a function of the off-wall velocity at $y^+ \approx 10$ and (b) Wall shear stress as a function of the off-wall velocity at $y^+ \approx 100$.}
	\label{fig:FNN_yplus=10_and_100}
\end{figure}

\section*{References}
\bibliography{main}
\end{document}